# Extracting morphometric information from rat sciatic nerve using optical coherence tomography

J. Hope[1,2], B. Bräuer[1], S. Amirapu[3], A. McDaid[2], F. Vanholsbeeck[1]

[1]*Dodd Walls Centre, The Department of physics, The University of Auckland, 38 Princes St, Auckland, NZ*
[2]*The Department of Mechanical Engineering, The University of Auckland, 5 Grafton Rd, Auckland, NZ*
[3] *Anatomy and Medical Imaging, University of Auckland, 85 Park Rd, Auckland, NZ*

## Abstract

We apply three optical coherence tomography (OCT) image analysis techniques to extract morphometric information from OCT images obtained on peripheral nerves of rat. The accuracy of each technique is evaluated against histological measurements accurate to +/-1 µm. The three OCT techniques are: 1) average depth resolved profile (ADRP); 2) autoregressive spectral estimation (AR-SE); and, 3) correlation of the derivative spectral estimation (CoD-SE). We introduce a scanning window to the ADRP technique which provides transverse resolution, and improves epineurium thickness estimates - with the number of analysed images showing agreement with histology increasing from 2/10 to 5/10 (Kruskal-Wallis test, $\alpha = 0.05$). A new method of estimating epineurium thickness, using the AR-SE technique, showed agreement with histology in 6/10 analysed images (Kruskal-Wallis test, $\alpha = 0.05$). Using a tissue sample in which histology identified two fascicles with an estimated difference in mean fibre diameter of 4 µm, the AR-SE and CoD-SE techniques both correctly identified the fascicle with larger fibre diameter distribution but incorrectly estimated the magnitude of this difference as 0.5µm. The ability of OCT signal analysis techniques to extract accurate morphometric details from peripheral nerve is promising but restricted in depth by scattering in adipose and neural tissues.

## 1 Introduction

In the field of neural prosthetics, the performance of nerve cuffs for recording and stimulating bio-electric signals can be improved using physiologically accurate volume conductor models of nerves [1, 2]. The morphometric details required for such models are the size, number and location of fascicles, the thickness of ultra-structural tissue layers, and the spatial variations in fibre diameter distribution. It is common to use simplified morphometric details, such as in [2-4], however, the results are not transferrable to patients due to patient specific tissue morphology. Another approach is to use destructive imaging methods, such as light microscopy, to acquire micrometre resolution histological images of the nerve cross section at one location and then extrude this along the length dimension, such as in [5, 6], however, this does not account for the length variation in tissue morphology caused by fascicle bifurcation [7]. Magnetic resonance imaging (MRI) enhanced with gadolinium-DTPA (diethylenetriamine penta-acetic acid) contrast agent, a non-destructive imaging method, has been used to image the size, number and location of fascicles in an extracted nerve tissue with a voxel size of 30 x 30 x 250 µm³ [1], which is promising, particularly if replicated with *in-vivo* measurements. For a patient-specific and physiologically accurate model, a non-destructive volumetric imaging method is required with a resolution of several µm.

The structure of peripheral nerves comprises one or more fascicles bound together by epineurium tissue 10's of µm thick [8]. In humans, the median nerve is several mm across and can contain 10 or more fascicles at the elbow, with each fascicle ranging in size from 0.12 to 2 mm² [9], whereas, in comparison, the rat sciatic nerve is approximately 1 mm across and can contain 3 to 4 fascicles ranging in size from 0.05 to 1mm². Each fascicle contains several thousand nerve fibres bound together by endoneurium tissue, and encompassed by a layer of perineurium tissue several µm thick [10]. Nerve fibres are long cylinders ranging in size from 1 µm to 22 µm diameter and are heterogeneously distributed within fascicles [9]; they are also highly aligned, densely packed, and usually sheathed in lipid rich myelin from encasing Schwann cells. The orders of magnitude of the dimensions of the nerves under study place Optical Coherence Tomography (OCT), a non-destructive imaging method, well as a potential means to acquire morphometric details without damaging the nerve.

Qualitative OCT techniques of distinguishing neural tissue from surrounding tissue [11-14], identifying different neural tissue layers [11-13, 15, 16], and analysing levels of myelination [17] do not provide quantified values nor confidence levels. On the other hand, quantitative OCT techniques, such as the depth-resolved analysis of optical properties [18] and statistical analysis of spectra [19, 20], provide quantified values for nerve tissue morphometry but have not been validated. Other quantitative OCT techniques, such as analysis of Mie scatter spectra [21] and optical scattering properties [22], have been used to classify tissue but have not yet been applied to peripheral nerves. There is therefore a need to evaluate and validate the performance of OCT techniques in imaging peripheral nerves, which builds on preliminary work in ref [23].

In this paper, we present results from three quantitative OCT signal analysis techniques that we identified in the literature and replicated with some improvements on images of rat sciatic nerve acquired with a swept source OCT (SS-OCT) system. Of the three OCT techniques - average depth resolved profile (ADRP), autoregressive spectral estimation (AR-SE), and correlation of the derivative spectral estimation (CoD-SE) – one, the former, was developed specifically for estimation of epineurium thickness while the latter two were developed for scatter size estimation applications. With each technique, we attempt two tasks: 1) extract the epineurium layer thickness, and 2) distinguish adipose tissue. In addition to this, for the two techniques based on scatter size estimation we attempt a third task: 3) estimation of fibre diameter distribution. We compare our results to histological analysis performed on light-microscopy images, a step which is absent in the original reporting of the OCT techniques [18-21, 24]. When replicating each technique, parameters were selected using unbiased methods, or otherwise noted as biased, to ensure fair comparison and practical application. We demonstrate new applications of the two scatterer size estimation techniques by using them to evaluate the combined thickness of epineurium and perineurium tissue, and to differentiate adipose tissue from neural tissues. The scatterer size estimation techniques were also used to evaluate the fibre diameter distribution of nerve fibres within fascicles. The results help establish the abilities and limitations of current OCT techniques in extracting morphometric details from peripheral nerves, and demonstrate the potential for OCT based scatterer diameter estimation techniques in this area.

## 2 Materials and Methods

### 2.1 Tissue preparation and handling

The animal procedures were approved by the University of Auckland Animal Ethics Advisory Committee. All animal specimens were rats of Wistar breed and male. Euthanasia was performed by first anaesthetising with isoflurane and then performing cervical dislocation. A total of 3 nerve tissue samples were explanted, each extending the entire length of the sciatic nerve and, distal to bifurcation, as much of the tibial and peroneal branches as practicable. Samples were then stored in 0.01M phosphate buffered solution at 4ºC for up to three days before imaging on OCT. Several points along each tissue sample were randomly selected and marked with Davidson's marking dye and 1% picric acid, and these points imaged with OCT. In this study we present results and images from one point each from the sciatic section of two nerves, and one point from the tibial section of one nerve.

During OCT imaging, tissue samples were suspended at two points 20mm apart along the length. Sagging of the tissue sample was used as a visual indication that the tissue sample was not mechanically stretched. Four OCT images, at 90º offsets, were acquired of the dyed points within the suspended section. Between image acquisitions 0.01M phosphate buffered solution was applied to the outside of the nerve, using a syringe, to avoid tissue drying. Post imaging, the tissue samples were returned to the phosphate buffered solution and transported to the Anatomy and Medical Imaging department at the University of Auckland for histological analysis by light microscopy.

Error introduced by thermal expansion, caused by variation in the sample temperature during imaging between 4ºC and 20ºC, is expected to be <0.5% using the thermal expansion coefficient of water. Error introduced from stretching, due to suspension of the tissue samples during OCT imaging, is <0.05% using a Poisson's ratio and Young's modulus of 0.37 and 41MPa respectively [25].

### 2.2 Microsphere samples

Three microsphere samples (Spherotech Inc.) were used in this study with concentration, and mean diameter +/- 1 standard deviation of: 1) 5% w/v, 3.8 +/-0.25 µm; 2) 5% w/v, 5.33 +/-0.25 µm; and 3) 2.5% w/v 8.49 +/-0.25 µm. Standard deviations of their respective size distribution were calculated using measured data provided by the supplier with the samples. For each sample, the vial containing the microspheres in solution was shaken vigorously to ensure homogeneity of the solution before a 1 mL sample was extracted and transported, using a pipette, to a separate container under the OCT system and immediately imaged.

**2.3** OCT image acquisition and processing

The OCT system has a swept source centred at 1310 nm with an 80nm bandwidth (3dB). The system has 12.5 µm axial resolution, 20 µm lateral resolution, and the 6 dB drop off in air is over 12mm. More details of the OCT system can be found in refs [26, 27].

The B-scan direction was perpendicular to the length of the nerve tissue samples, spanned a physical distance of 5mm, and contained 714 A-scans. B-scans were saved from an LABVIEW® user interface to text files and then processed individually with MATLAB® 2015b using the signal analysis techniques described below.

In order to determine the distance when calculating epineurium thickness, an average refractive index had to be used for the entire OCT image; here, we have used the refractive index of n=1.40, which is the average of values reported for myelin (1.455) [28] and bovine tendon, a collagenous connective tissue, (1.353) [29]. The axial distance per pixel, of 10 µm in air, was, therefore, 7.1 µm in neural tissues in all subsequent calculations of epineurium thickness.

**2.4** Histological analysis by light microscopy

Histology by light microscopy was performed at 1 location on each of the tissue samples, see Fig. 1. Tissue samples were preserved in 10% Neutral Buffered formalin for 24 hours, followed by 70% ethanol prior to paraffin embedding. Tissue slices 10 µm thick were obtained using a microtome then Haemotoxylin & Eosin was used to stain the collagen rich epineurium, perineurium and endoneurium tissues pink. Lipid rich myelin was stained with Luxol Fast Blue. Images were obtained on a Leica DM500 light microscope at 4x, 10x and 40x magnifications. A uniform shrinkage of 10%, produced during the preparation process, was taken into account by multiplying all values extracted from histology by a factor of 1.1.

The thickness of epineurium tissues was determined using images obtained at 4x magnification. In each image, a line was drawn through the tissue layers, perpendicular to the tissue outer surface, at 20 locations approximately equidistant around the fascicle boundaries. Then, the thickness of tissue along these lines was measured manually. Error introduced by the measurement procedure was estimated to be +/- 1 µm.

Fibre diameter distribution was calculated in areas of the tissue sample images which were visually identified as containing few artefacts from sample preparation. A grid of 100 µm squares was laid over the area of interest in the 40x magnification images and the fibre diameter distribution was calculated for each grid square. Fibres were typically ellipsoid shaped, therefore, the major and minor axes were used to calculate the diameter of a circle with equivalent area. The diameter estimates contained approximately +/-1 µm uncertainty. Furthermore, the method employed could not identify fibre diameters less than 3 µm due to the very thin, or absent in the case of unmyelinated fibres, myelin sheath.

**2.5** Average depth resolved profile (ADRP) techniques

Two techniques to identify the fascicle boundary, one using structural data and the other phase retardation data, are presented in ref [18]. Both techniques are applied to the average depth-resolved profile (ADRP) of the middle 50% of A-scans of the tissue sample. The ADRP is generated by first cropping pixels outside the tissue sample, then flattening the remaining pixels so that each row in the columns of data (A-scan) correlates to the same axial depth in the tissue sample, and, finally, averaging values across each row. In order to estimate the fascicle boundary with higher lateral resolution we implemented the ADRP across windows containing 5 A-scans, which were scanned laterally across the sample in 1 pixel steps.

To identify the fascicle boundary the ADRP-structural-data technique plots the signal to noise ratio (SNR) of the ADRP on a logarithmic scale against depth, then fits a linear slope to the uniformly decaying region of the curve, Fig 1. The absolute difference between the linear fit and the ADRP curve forms the residual SNR curve. A threshold value is then calculated as the mean plus 2 standard deviations from the linear portion of the residual SNR curve. The fascicle boundary is identified by the intersection of the threshold value and the curve neighbouring the uniformly decaying region. In our implementation of the 5 A-scan sliding window method, we fit the linear slope to data from a depth range of 15 to 50 pixels in all window ADRPs, Fig 1. The start value of 15 pixels, correlates to a depth of 113 µm, which is outside the typical epineurium thickness values provided by histological analysis and, therefore, is highly unlikely to include epineurium tissue.

To identify adipose tissue, we calculated the threshold from the entire portion of the residual SNR curve, instead of just the linear portion, which increases the threshold value. In areas with epineurium, the residual-SNR curve in the uniformly decaying region was below the threshold and the initial peaks of the residual-SNR curve, associated with epineurium, was above the threshold. Conversely, in areas with adipose tissue, which contained multiple peaks and no uniformly decaying region, the residual SNR curve oscillated around the threshold.

In the ADRP-phase-retardation technique, a linear slope is fitted to the rising portion in a plot of phase retardation against depth. The residuals, threshold and fascicle boundary are then calculated in the same way as the ADRP-structural-data technique. We did not employ the ADRP phase-retardation technique on any tissue samples as our system was not polarisation sensitive.

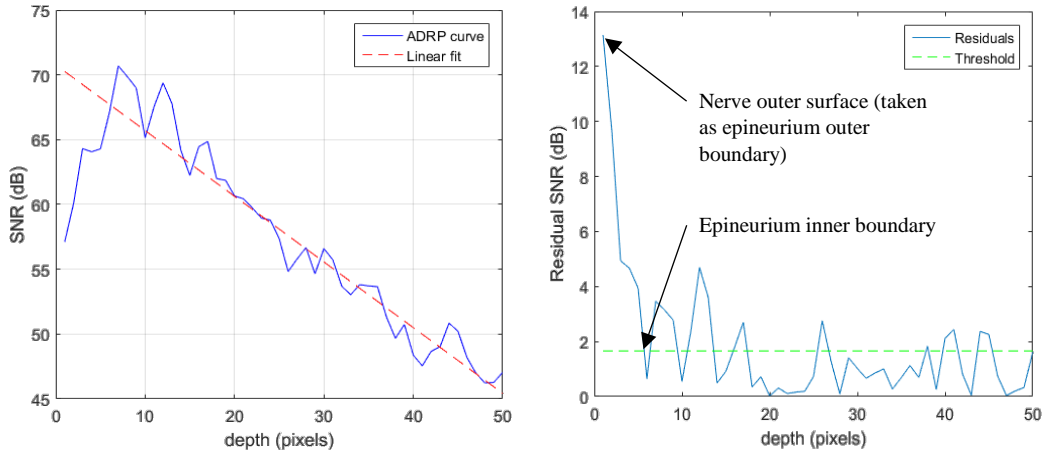

Figure 1: An ADRP curve generated by averaging 5 adjacent Ascans from a flattened OCT image and the linear fit calculated from pixel depth range of 15 to 50 pixels (a), where a pixel depth of 0 is the surface. The residuals, calculated from Fig 1a, and the Threshold line used to determine the epineurium inner boundary (b).

## 2.6 Autoregressive spectral estimation (AR-SE) technique

The AR-SE technique for estimating scatterer diameter was originally presented in ref [30] and then applied to biological samples, including the sciatic nerve of rabbit, in refs [19, 20]. The technique assumes a linear relationship between the scatterer diameter $\widetilde{\mathbf{d}}$ and the eigenvectors produced by principal component analysis of the power spectral density $\widetilde{\mathbf{P_{xx}}}$ produced using Burg's method, such that $\widetilde{\mathbf{P_{xx}}}\widetilde{\mathbf{A}} = \widetilde{\mathbf{d}}$ where $\widetilde{\mathbf{A}}$ is a coefficient matrix which is populated using samples of known diameter, also called 'training data'.

To obtain the training data, we used the three microsphere samples described above (section 2.2). A window 50 x 10 pixels (axial x transverse) in size was raster scanned across half of the OCT image at 25 and 1 pixel steps respectively. Data was row-averaged to suppress noise and normalised to reduce variations from depth attenuation. The decay of reflection coefficients towards zero determined the autoregressive order to use in calculating the power spectral density. A target variance of 99.99% determined the number of components to extract from weighted principal component analysis (PCA) of the power spectral density data. The remaining half of each OCT image of microspheres was used to obtain diameter estimates.

In applying the AR-SE technique to OCT data from tissue samples, larger window sizes reduced noise but potentially removes more of the targeted detail such as the intra-fascicle boundary and localised variation in fibre diameter distribution. Furthermore, the maximum autoregressive order that can be applied is one less than the number of data points, i.e. the axial window size. A window size of 10 x 10 pixels, scanned at 1 pixel steps, was selected as it produced a good balance between noise reduction and sufficient resolution. At each step, the autoregressive order to use was determined by analysing the reflection coefficients, with a drop below +/-0.1 in magnitude used as the threshold criterion. Alternatively, when the decay of reflection coefficients towards zero was not clearly identifiable, an autoregressive order of 3 was selected as this value was common in the tissue samples and was found to minimise noise in the solution.

**2.7** Correlation of the derivative spectral estimation (CoD-SE) technique

The CoD-SE technique was originally presented in ref [24], and then again with additional signal processing steps in ref [21]. This technique uses Mie theory to generate spectra for spherical scatterers before taking the derivative and then the autocorrelation to produce the correlation of the derivative (CoD). The CoD bandwidth is calculated as the first minimum, i.e. the minimum with the lowest lag value, in the CoD of the spectra. A curve is fitted to a plot of the theoretical relationship between the scatterer diameter and CoD bandwidth. In OCT data, the spectra are obtained using the Fourier transform with a Gaussian window. In order to lower the noise, spectra are low pass filtered and the edges are removed. Differentiating with respect to the neighbouring value followed by autocorrelation produces the CoD. CoD-SE is highly sensitive to the lateral position of the window relative to the scatterer, and to the axial window size. To address the former, an intensity threshold of 5dB above the noise level is introduced and the highest intensity of 3 laterally adjacent windows is assigned to all 3 lateral positions. To address the latter, the axial window size, which is used to obtain sample spectra, is selected to minimise the standard deviation of the resulting scatterer diameter estimates.

We appended an additional step to the end of the original method, outlined in the previous paragraph, in order improve the scatterer diameter estimates produced from our microsphere samples: the window axial size was selected to minimise the 'normalised' standard deviation of the scatterer diameter estimate, calculated as the standard deviation divided by the corresponding mean.

The CoD-SE algorithm was tested on three microsphere samples with same concentration and mean diameter as in the AR-SE experiments, see section 2.5. We generated the theoretical curve using the MATLAB functions of Mie theory presented in refs [31, 32] and refractive indices of water and polystyrene [33] ($n_{water} = 1.3225 + 0.001i$ and $n_{poly} = 1.59 + 0.0025i$). We then fitted an exponential function of the form $f_{CoD} = d^A e^B$ where $f_{CoD}$ is the CoD bandwidth, $d$ is diameter, and $A$ & $B$ are constants defining the curve, as this provided an excellent fit within the diameter range 3 µm to 16 µm. Unbiased methods were not identified to select 1) the standard deviation of the Gaussian window in the Fourier transform, and 2) the extent to crop the edges of spectra to remove noise. A standard deviation of the Gaussian window equal to half of the window size, and cropping of 3.9 nm (100/4096 points) from each end of the spectral range, were selected through trial and error using the microsphere samples. Diameter estimates were produced for each image pixel using 13 different axial window sizes from 3 to 15 pixels, all with a lateral size of 1 pixel, scanned at 1 pixel steps. At each step, and for each window size, the Butterworth filter ($2^{nd}$ order, zero phase, low pass) cut-off frequency was selected using Winter's method, with resulting values in the range 0.225 +/- 0.004 π.rad/sample (0.899 +/-0.016 nm⁻¹).

For tissue samples, we could not find a value for the refractive index of endoneurium nor its close relative epineurium. Instead, we used bovine tendon, which is, like endoneurium, a collagen rich connective tissue. Therefore, we generated the theoretical CoD bandwidth curve using refractive indices of myelin ($n_{myelin} = 1.455$ [28]) and bovine tendon ($n_{tendon} = 1.353$ [29]), to which we fitted an exponential function of the form $f_{CoD} = d^A e^B$. Values for the standard deviation of the Gaussian window in the Fourier transform, the extent to crop the edges of spectra, the Butterworth filter parameters were all carried over from microsphere experiments. Axial window sizes from 3 to 15 pixels were used to generate diameter estimates, and then a 3 x 3 scanning window was used to determine the diameter estimate to use for each pixel location based on minimisation of the normalised standard deviation. Scatterer size estimates were directly assigned as fibre diameter distribution estimates.

# 3 Results

## 3.1 Epineurium thickness

### 3.1.1 Histological analysis

With H&E staining the fascicle, boundaries in the sciatic branches were clearly identifiable in light microscopy images of all three tissue samples by the change from dark pink stained epineurium and perineurium tissues to the lighter pink intra-fascicle environment mottled with unstained nerve fibres, Fig. 2. Adipose cells are unstained by H&E. Adipose tissue commonly appeared in clusters of several cells, each with a diameter of >20 µm, apposed to the nerve outer boundary was present in large quantities in tissue sample 1 and in small quantities in tissue samples 2 and 3, Fig 2.

The combined epineurium and perineurium thickness was estimated as 50 +/- 38 µm, 45 +/- 32 µm and 31 +/- 25 µm (mean +/- 1 standard deviation) across the entire light microscopy image for tissue samples 1, 2 and 3 respectively.

Due to the small thickness of perineurium relative to epineurium, and because one layer of each are always present and adjacent to one another in the tissue layer separating the intra-fascicle volume and the outer nerve boundary, from here onwards we refer to the combined epineurium and perineurium tissue layer as just the epineurium layer.

When the light microscopy images were aligned to the corresponding OCT image, and the middle 50% of the nerve boundary was analysed to match the area imaged and analysed by OCT, the localised standard deviations of the epineurium thickness estimates were smaller than those obtained for the entire nerve. This indicates epineurium thickness varies along the outer boundary. The distribution of the thickness estimates for tissue sample 1 orientation 3 and tissue sample 3 orientations 1 & 2 showed multiple peaks across the range of values, i.e. multimodal. All other tissue samples' orientations were right skewed, i.e. favouring smaller thickness, see Table 1.

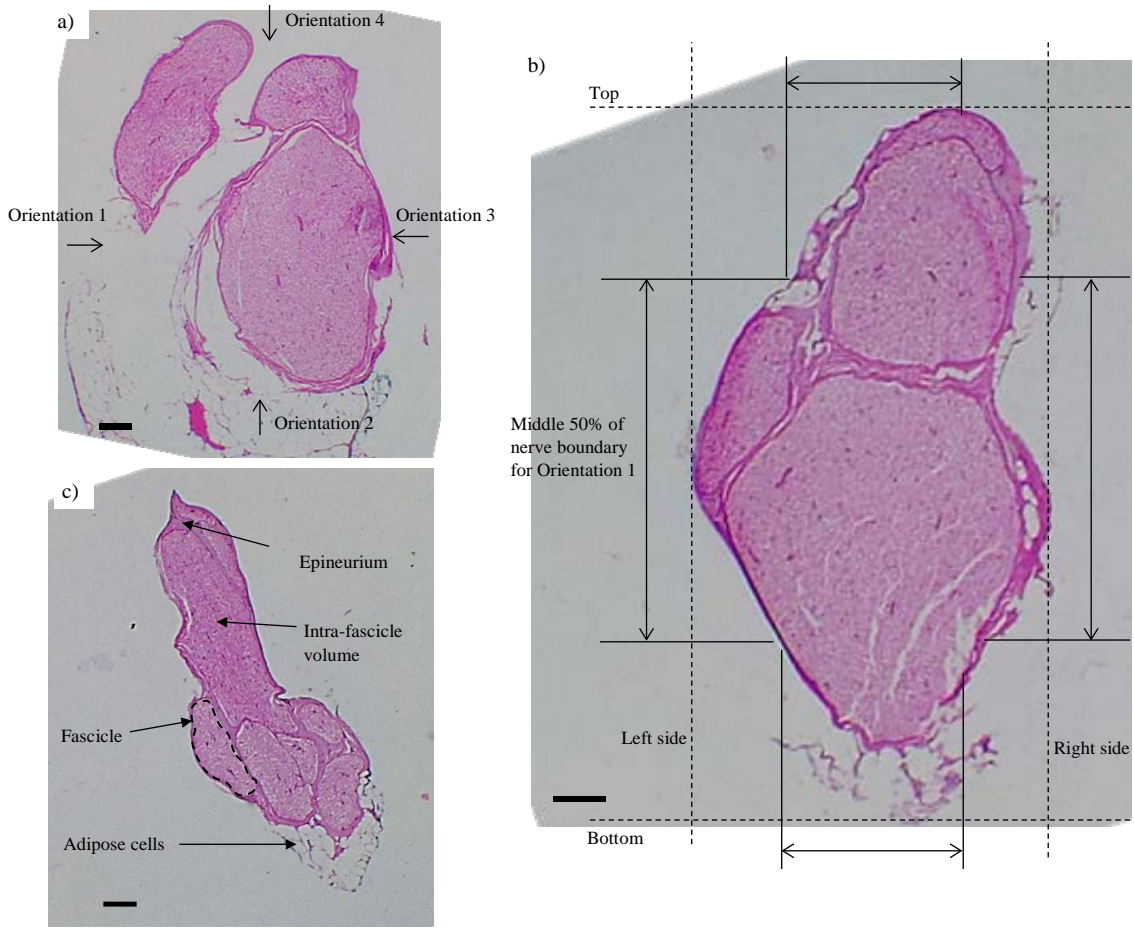

Figure 2: Light microscopy images at 4x magnification, stained with H&E, of tissue sample 1 (a) and tissue sample 2 (b) of sciatic nerves, and tissue sample 3 (c) of the tibial nerve branch. In all images the scale bar is 100 µm. Images have been rotated in order that the incident beam in OCT, acquired for four tissue orientations, coincides with each side of the light microscopy image, where the sides: left, bottom, right, top correspond to OCT imaging orientations: 1, 2, 3, 4 respectively (a). After alignment with OCT, the top, bottom and two sides of the nerve within each histological image were identified, shown with dashed lines in (b), and the middle 50% of the nerve boundary was assigned as the middle 50% in linear distance between the respective ends, shown with arrows in (b). Tissue features of peripheral nerve are labelled in (c).

### 3.1.2 Average depth resolved profile (ADRP) techniques

Using the conventional ADRP-structural-data method, to construct the ADRP from the middle 50% of A-scans, estimates of the combined epineurium thickness from all orientations of tissue samples 1 and 2 were consistently lower than results obtained from sciatic nerve presented in the original study for the ADRP-structural-data technique of 133 +/-14 µm [18]. Tissue sample 3 is from the tibial nerve branch and so is not directly

Table 1: Epineurium tissue layer thickness, in µm, estimated from histology, conventional and 5Ascan methods within the ADRP technique, and with a 10 x10 window for the AR-SE technique. Mean (standard deviation). A refractive index of 1.41, producing 7.1 µm axial distance per pixel within nerve tissue, used to obtain ADRP and AR-SE values. Histology estimates are taken across the middle 50% of each orientation, whereas the ADRP and AR-SE signal analysis techniques are applied to the middle 50% of A-scans in each image. 'RS' = Right skewed, 'MM' = Multimodal. In the ADRP conventional column, blue shading indicates values within +/-1 standard deviation of the value obtained using light microscopy. Elsewhere, green shading indicates agreement of medians (Wilcoxen rank sum test, $\alpha = 0.05$) and analysis of variance (Kruskal-Wallis test, $\alpha = 0.05$) with the corresponding data obtained using light microscopy.

| | Light microscopy | ADRP: Conventional | ADRP: 5Ascan | AR-SE: 10x10 |
|---|---|---|---|---|
| Tissue sample 1 | | | | |
| orientation 1 | * | 77 | 40 (37) | 70 (75) |
| orientation 2 | * | 91 | 42 (39) | 72 (50) |
| orientation 3 | 59 (34) MM | 63 | 48 (28) RS | 27 (24) RS |
| orientation 4 | 19 (10) RS | 85 | 37 (27) RS | 17 (22) RS |
| Tissue sample 2 | | | | |
| orientation 1 | 28 (19) RS | 83 | 44 (28) RS | 43 (50) RS |
| orientation 2 | 29 (11) RS | 91 | 61 (32) MM | 22 (23) RS |
| orientation 3 | 44 (20) RS | 91 | 56 (34) MM | 44 (41) RS |
| orientation 4 | 30 (25) RS | 105 | 33 (34) RS | 26 (31) RS |
| Tissue sample 3 | | | | |
| orientation 1 | 23 (14) MM | 77 | 57 (38) RS | 43 (43) RS |
| orientation 2 | 41 (29) MM | 46 | 28 (29) RS | 28 (39) RS |
| orientation 3 | 22 (12) RS | 71 | 40 (37) RS | 20 (25) RS |
| orientation 4 | 22 (16) RS | 49 | 51 (35) RS | 27 (32) RS |

* adipose tissue prevalent

comparable. Epineurium thickness estimates obtained using the conventional ADRP-structural-data technique on tissue sample 1 orientation 3 and tissue sample 2 orientation 2 are within +1 standard deviation, and tissue sample 3 orientation 4 within +2 standard deviations of the mean values provided by measurement from corresponding histology by light microscopy. The remaining estimates are greater than + 2 standard deviations, indicating significant disagreement between values, Table 1.

Thickness estimates using the 5 A-scan sliding window method were consistently lower than those obtained using the conventional ADRP method. Furthermore, with the exception of tissue sample 2 orientations 2 and 3, the distribution of thickness estimates were right skewed, i.e. favouring smaller thickness. With the exception of Tissue sample 2 orientation 2, all estimates in Table 1 obtained using the ADRP-5Ascan method are within 1 standard deviation of the values obtained from light microscopy, however, the right skewed distributions of data make means a poor measure for comparison. The medians (Wilcoxen rank sum test, $\alpha = 0.05$) and analysis of variance (Kruskal-Wallis test, $\alpha = 0.05$) of estimates obtained with the ADRP-5Ascan from tissue sample 1 orientation 3, tissue sample 2 orientations 3 & 4, and tissue sample 3 orientations 2 & 3 were in agreement with the corresponding histology by light microscopy.

### 3.1.3 Autoregressive spectral estimation (AR-SE) technique

The layer of epineurium and perineurium tissues was visually identified as a contiguous area of large diameter scatterers along the surface of the nerve. The relationship between the thickness of this contiguous layer of scatterers and the thickness of the epineurium tissue was investigated using a digital phantom of an epineurium layer, which was constructed using a 50x50 element array with each element equal to 0.01 and within this array several adjacent rows containing values of 1 to mimic epineurium. Application of the AR-SE algorithm to the phantom showed the epineurium thickness along each A-scan could be approximated by difference between the base width of the contiguous scatterer peak and the window axial size. In nerve samples, due to the presence of scatterers along the inside boundary of the epineurium tissue layer, we isolated the contiguous scatterer peak by removing scatterer estimates below 5 µm. We then subtracted the scanning window axial size from the full-width-half-mean of the contiguous scatterer peak, in place of the base width, to estimate the epineurium thickness.

Visual identification of the layer epineurium tissue, as a contiguous area of large diameter scatterers along the surface of the nerve, is an observation not reported in the original study. The distributions of thickness estimates produced with the AR-SE technique are generally right skewed with a long right-hand tail which makes the means a poor measure for comparison, Fig. 3. Indeed, none of the AR-SE thickness estimate means were in agreement with their corresponding histology by light microscopy (two sample Kolmogorov-Smirnov test, α = 0.05). Estimate medians (Wilcoxen rank sum test, α = 0.05) and analysis of variance (Kruskal-Wallis test, α = 0.05) of tissue sample 1 orientation 4, all orientations of tissue sample 2, and tissue sample 3 orientation 4 were in agreement with their corresponding histology by light microscopy.

It is evident in box and whisker plots of AR-SE estimates in Fig. 3 that in all tissue samples and orientations there was at least one, and typically several, thickness estimates of 0 µm, which is biologically impossible.

### 3.1.4 Correlation of the derivative spectral estimation (CoD-SE) technique

Scatterer diameter estimates produced by the CoD-SE algorithm varied depending on the size of the square scanning window, with larger square windows favouring smaller mean diameter estimates with smaller standard deviations. An unbiased method of selecting this parameter is desirable for practical application of the CoD algorithm. On OCT data of nerve tissue, a contiguous layer of estimates within the range of 2.2 to 4 µm were produced around the tissue boundary where the epineurium layer is expected to reside. Although this narrow diameter range may potentially provide a means to distinguish the epineurium layer from the adjacent air on one side and intra-fascicle volume on the opposing side, we did not identify a method to easily isolate this layer from scatterers present along its internal and external boundaries. In addition, we did not extract a relationship between this contiguous layer and the epineurium thickness through application of the CoD-SE algorithm to a digital phantom of an epineurium layer. Therefore, we did not pursue epineurium thickness estimation with the CoD-SE technique further.

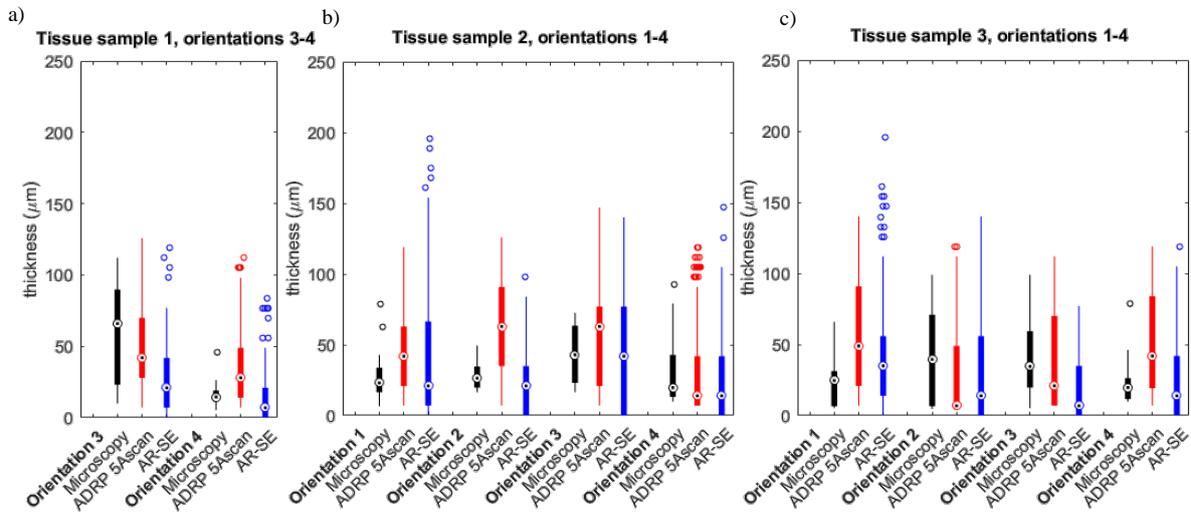

Figure 3: Distribution of epineurium tissue layer thickness estimates for tissue samples 1 (a), 2 (b) and 3 (c). Within each tissue sample boxplot, estimates from histology by microscopy (black), the 5 A-scan window method within the ADRP technique (red), and AR-SE technique with a 10 x 10 window (blue), are grouped together for each orientation.

## 3.2 Fibre diameter distribution

### 3.2.1 Histological Analysis

Tissue sample 1 orientation 4 contained easily identifiable landmarks – in the form of two 'humps' – for alignment to corresponding OCT images, and contained two fascicles with a difference in fibre diameter distributions that was easily perceptible at 10x and 40x magnifications, Fig. 4a. Tissue sample 1 orientation 4 was therefore selected for fibre diameter distribution analysis. A minor histological artefact is evident in this sample where the left fascicle has been pulled away from the right fascicle during slicing on the microtome, Box and whisker plots of the fibre diameter distributions within the left (grids 1 – 9) and right (grids 10 – 18) fascicles showed higher medians, upper and lower quartiles, and maximum values in the majority of grids in the right fascicle compared to grids of the left fascicle, Fig. 4b. Further analysis, using one-way ANOVA with $\alpha = 0.05$, revealed significant difference between the means of grids 1 and 3 within the left fascicle, and no significant difference between the means of grids in the right fascicle. The left and right fascicles possessed diameters of 6.1 +/-2.1 µm and 10.1 +/- 3.0 µm (mean +/- 1 standard deviation) respectively. One-way ANOVA, $\alpha = 0.05$, also revealed significant difference in fibre diameter between the two fascicles, indicating a heterogeneous fibre diameter distribution at the fascicle level.

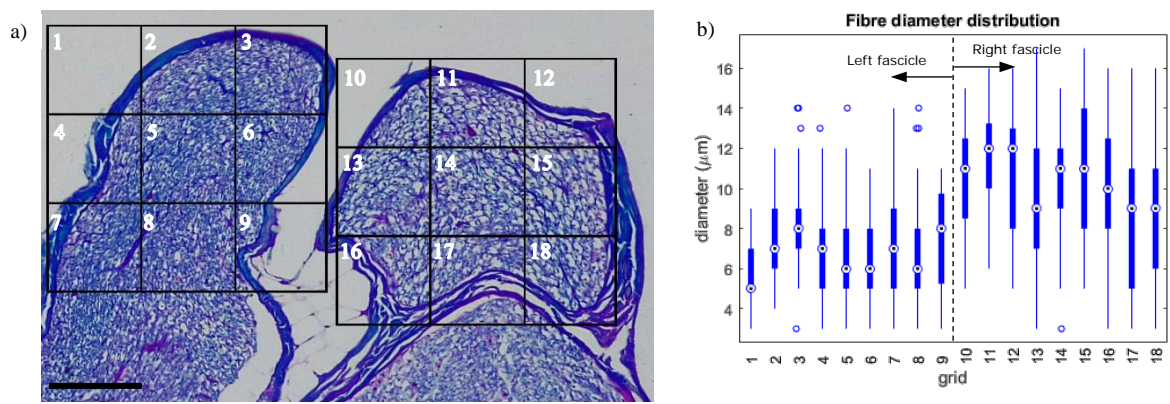

Figure 4: Light microscopy image of histology at 10x magnification of tissue sample stained with LF Blue (a), scale bar is 100 µm. Numbered grid used in fibre diameter distribution calculations are shown overlaid on the light microscopy image (a), with results presented in box and whisker plots (b) confirming significant difference in fibre diameter distribution between the left (grids 1 – 9) and right (grids 10 – 18) fascicles.

Table 2: Scatterer diameter estimates, in µm, of microsphere samples and the intra-fascicle volume of the left and right fascicles in tissue sample 1 orientation 4. Estimates from histology, a 10 x10 window in the AR-SE technique, the 'CoD-SE: conventional' technique - which selects each estimate based on minimisation of the standard deviation, and the 'CoD: normalised s.d. technique – which selects each estimate based on minimisation of the normalised standard deviation. Data are presented as mean (standard deviation) in µm.

| | Manufacturer's data / Histology [µm] | AR-SE [µm] | CoD-SE: conventional [µm] | CoD-SE: normalised s.d. [µm] |
|---|---|---|---|---|
| 3.8 µm microspheres | 3.8 (0.25) | 3.6 (0.9) | 4.3 (1.5) | 4.3 (1.5) |
| 5.33 µm microspheres | 5.33 (0.25) | 5.0 (1.2) | 4.6 (1.5) | 5.7 (1.4) |
| 8.49 µm microspheres | 8.49 (0.25) | 7.9 (2.1) | 6.3 (1.9) | 6.3 (1.9) |
| Tissue sample 1 orientation 4 | | | | |
| Left fascicle | 6.9 (2.2) | 7.6 (4.8) | 9.7 (3.3) | 10.4 (4.4) |
| Right fascicle | 10 (3.1) | 8.1 (4.6) | 9.6 (3.2) | 10.9 (4.4) |

### 3.2.2 Autoregressive spectral estimation (AR-SE) technique

The AR-SE technique estimated the diameters of microsphere samples of 3.8, 5.33, and 8.49 µm diameter as 3.6 +/- 0.90, 5.0 +/- 1.2, and 7.9 +/- 2.1 µm (mean +/- 1 standard deviation) respectively, Table 2. These results show a good diameter estimate accuracy (t-test, $\alpha = 0.05$) but with high probability of large (>25%) errors in individual measurements.

On OCT data from tissue sample 1 orientation 4, the AR-SE technique predicted a heterogeneous distribution of fibre diameters in both fascicles, Fig. 5b. A small difference between the scatterer diameter distributions in the left and right fascicles was predicted, with means of 7.6 and 8.1 µm respectively, table 2, and medians of 8.2 and 9.0 respectively, Fig 5c. The diameter distribution estimates were significantly different to those produced from the corresponding histology by light microscopy images (t-test, α = 0.05; and Wilcoxen rank sum test, α = 0.05), Fig. 5c. Therefore, in the example analysis of this image, the AR-SE technique correctly predicted a larger scatterer diameter distribution in the right fascicle but did not accurately quantify the difference between the two fascicles' distributions.

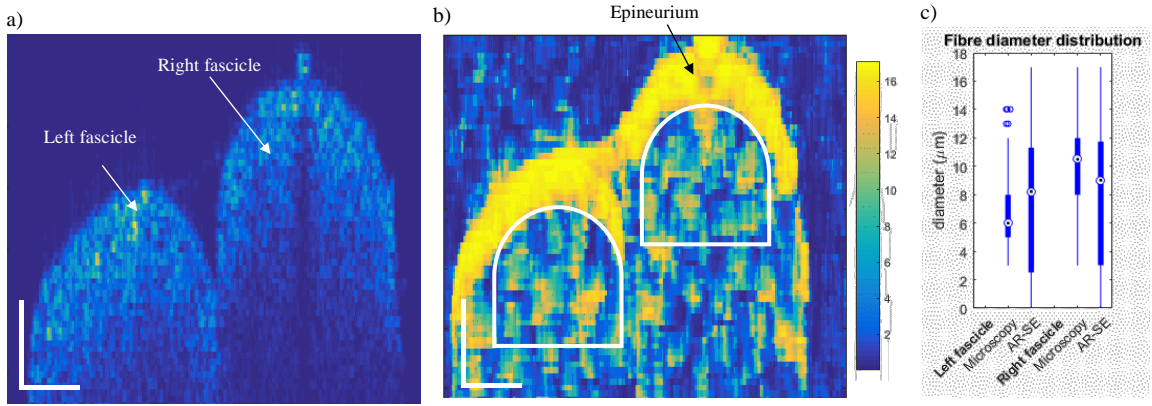

Figure 5: OCT data of tissue sample 1 orientation 4 (a) and the same image after processing with AR-SE technique (b), scale bars in (a) & (b) are 100 µm. In (b) the epineurium layer is visible as a contiguous layer of large diameter scatterers, with diameter on the colour scale. The white outlined region in (b) indicate the areas used to generate the fibre diameter distribution estimate with the CoD-SE algorithm for each fascicle, which are compared against the results from histology by light microscopy in a box and whisker plot (c). The AR-SE technique correctly predicted a larger scatterer diameter distribution in the right fascicle but did not accurately quantify the difference between two fascicles' distributions.

### 3.2.3 Correlation of the derivative spectral estimation (CoD-SE) technique

The CoD technique estimated the diameters of microsphere samples of 3.8, 5.33, and 8.49 µm diameter as 2.6 +/- 0.7, 2.8 +/- 0.5, and 4.5 +/- 0.3 µm (mean +/- 1 standard deviation) respectively, which indicates poor diameter estimate accuracy for all microsphere diameters, Table 2. All estimates corresponded to a scanning window axial size of 4 pixels, selected to minimise the standard deviation, whereas the smallest window axial size used in the original study was 5 pixels [21]. Excluding window axial sizes of 3 pixels and 4 pixels from consideration provided a significant improvement in estimate accuracy: 4.3 +/- 1.5, 4.6 +/- 1.5, and 6.3 +/- 1.9 µm respectively. For all microsphere samples, the magnitude of diameter estimates showed a decreasing trend with decreasing axial window size, as did the standard deviation of estimates, which results in smaller diameter estimates being favoured. Selection via minimisation of the normalised standard deviation removed the effect of the downwards trends, and, with window axial sizes of 3 pixels and 4 pixels excluded, increased the estimate for 5.33 µm microspheres to 5.7 +/- 1.4 µm. Estimates for 3.8 and 8.49 µm microspheres remained unchanged, Table 2.

On OCT data of nerve tissue, pixels containing air produced an estimate of 1.775 µm, which corresponded to a lag of 1500(a.u.) and was the limit of the power fit to the theoretical CoD curve. As mentioned earlier, in section 3.2.4, epineurium produced estimates within the range of 2.2 to 4 µm. Therefore, to remove air and epineurium, scatterer diameter estimates of less than or equal to 4 µm were discarded in our analysis of fibre diameter distribution. On OCT data from tissue sample 1 orientation 4, the CoD-SE algorithm predicted a heterogeneous distribution of fibre diameters in both the left and right fascicles, Fig. 6b. A small difference between the scatterer diameter distributions in the left and right fascicles was predicted, with means of 10.4 and 10.9 µm respectively, table 2, and medians of 10.0 and 10.6 µm respectively, Fig 6c. As for the AR-SE technique, the normalised CoD-SE technique correctly predicted a larger mean scatterer diameter in the right fascicle, but did not accurately quantify the difference.

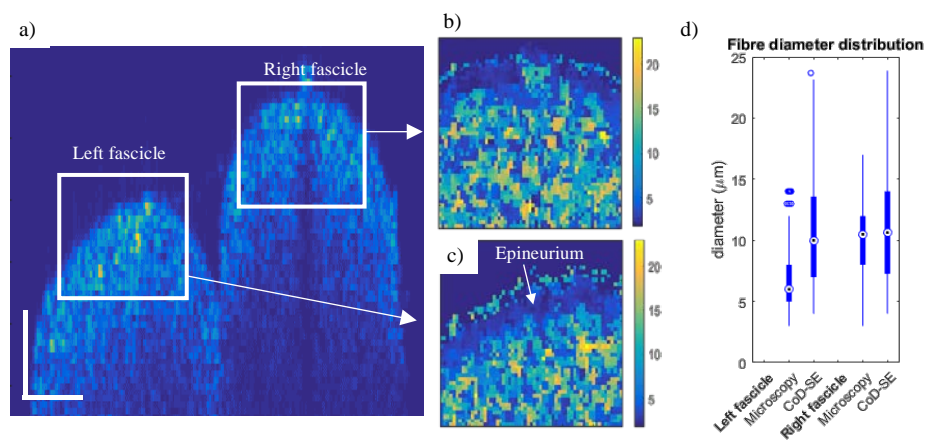

Figure 6: OCT data of tissue sample 1 orientation 4 (a) with white boxes indicating the areas of the left and right fascicles used for analysis with the CoD-SE algorithm, and the white scale bar bottom left indicating 100 µm. The boxed areas in (a) are shown after processing with CoD-SE algorithm in (b) and (c), for the right and left fascicles respectively, with diameter on a colour scale. The diameter distribution of areas (b) and (c), with values below 4 µm excluded, are compared against results from histology by light microscopy in a box and whisker plot (d). The normalised CoD-SE technique correctly predicted a larger mean scatterer diameter in the right fascicle, but did not accurately quantify the difference.

### 3.3 Adipose tissue identification

#### 3.3.1 Average depth resolved profile (ADRP) technique

Using the 5 A-scan sliding window method and calculating the threshold from the entire portion of the residual SNR curve, regions with adipose tissue apposed to the outer nerve boundary cross the threshold several times throughout the depth, Fig. 7b. In contrast, regions with epineurium crossed the threshold only once Fig. 7a. Thus, layers of adipose tissue were qualitatively distinguishable from epineurium tissue, a finding not reported in the original study [18]. Further studies on this effect are needed to establish the repeatability of this observation. It was not possible to identify the outer epineurium boundary beneath adipose tissue, and, as a result, the thickness of adipose tissue could not be estimated, nor was it possible to identify the uniformly decaying region indicating the intra fascicle volume beneath adipose tissue. The presence of adipose tissue, therefore, reduces the amount of information available from the ADRP technique.

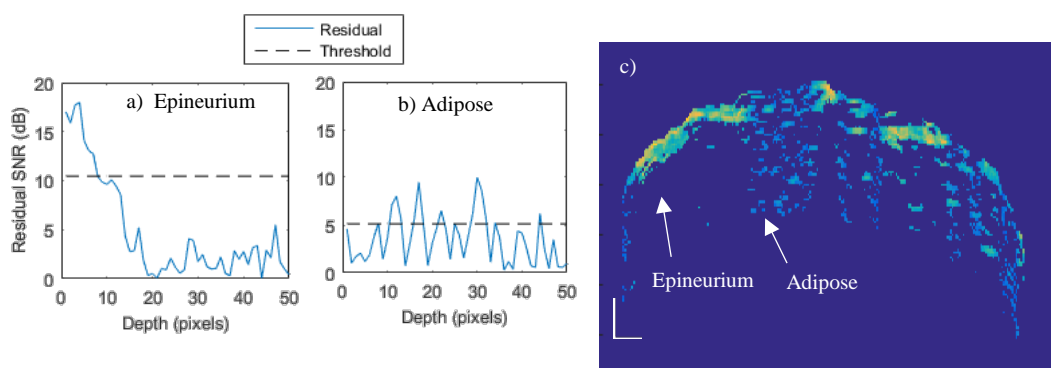

Figure 7: Residual SNR curve of epineurium (a) is clearly distinguishable from adipose tissue (b) as the threshold, when calculated from the entire residual curve, intercepts with a single large peak. A colour map of tissue sample 1 orientation 2 produced by the ADRP-5Ascan method showing only above-threshold residual SNR data (c) with white scale bar bottom left indicating 100 µm; a significant amount of adipose tissue is visible.

### 3.3.2 Autoregressive spectral estimation (AR-SE) technique

Adipose tissue was qualitatively identified within the AR-SE technique as clusters of large diameter scatterers present in areas of the image where the contiguous layer of scatterers on the nerve surface, characteristic of epineurium tissue, was absent. It was not possible to identify the boundary between epineurium and adjacent adipose tissue, and, as a result, the thickness of adipose tissue could not be estimated. A heterogeneous distribution of nerve fibres could feasibly also appear as clusters of large diameter scatterers. Thus, it was not possible to distinguish adipose tissue from the intra-fascicle volume. Similarities between AR-SE estimates of adipose tissue, epineurium, and heterogeneous fibre diameter distribution meant adipose tissue could not be identified with confidence without referencing against the corresponding light microscopy images.

### 3.3.3 Correlation of the derivative spectral estimation (CoD-SE) technique

Adipose tissue appeared as a heterogeneous distribution of scatterers within the range of 2 – 19 µm. A method of distinguishing adipose tissue from intra-fasicle volume was not found. The 50 µm and greater size of adipose cells, observed in histology by light microscopy, is beyond the Mie scattering range and so their true size cannot be estimated using the CoD-SE technique.

## 4 Discussion

A reference value for the combined thickness of epineurium and perineurium can be estimated by summation of the epineurium thickness range in ref [8] of 47 +/-25 µm for Sprague-Dawley rats weighing 350 to 450 g, and the perineurium thickness range from ref [10] of 3 to 5 µm for Wistar rats weighing 200 to 250 g, producing a combined thickness of approximately 51 +/- 26 µm. This reference value is in broad agreement with the values we obtained from histology, of 50 +/- 38 µm, 45 +/- 32 µm (mean +/- 1 standard deviation) for tissue samples 1 and 2 respectively. The value for tissue sample 3, of 31 +/- 25 µm, is not directly comparable as it is from the tibial, not the sciatic, branch. It is evident from the size of the standard deviations, which are large relative to mean values in both ref [8] and our estimates, that there is a significant variation in epineurium and perineurium tissue thickness across the cross section of each tissue sample.

When the ADRP was calculated using the middle 50% of A-scans, the technique produced estimates of epineurium thickness with an average of 86 µm, and which were consistently lower than those in the original paper, [18], which reported an average thickness of 133 µm on sciatic nerve of Sprague-Dawley rats. Both of these values, 86 µm and 133 µm, are significantly above our reference value of 51 +/- 26 µm [8, 10]. In addition, all of the epineurium thickness estimates we produced using this technique were consistently and significantly larger than the corresponding estimates using histology by light microscopy. The technique, therefore, appears to consistently and significantly overestimate the epineurium thickness. Accuracy of epineurium thickness estimates using the ADRP technique were significantly improved through the addition to the method of a scanning window containing 5 adjacent A-scans. The 5 Ascan ADRP technique also resolves the image with a higher transverse resolution, which is beneficial to practical application of the technique due to the variable epineurium thickness observed in the histology by light microscopy. A qualitative method of identifying adipose tissue through the residual SNR is a new application of the ADRP technique. However, further studies are required to evaluate the repeatability of the proposed method.

Use of the structural and optical differences within the peripheral nerve to estimate epineurium layer thickness is a novel application of scatterer diameter estimation techniques, such as AR-SE and CoD. In the AR-SE technique, our observation of epineurium as a contiguous layer of large diameter scatterers presents a new means to identify the epineurium boundary. Our initial results of thickness estimates, using the difference between the full-width-half-mean of the contiguous scatterer peak and the axial scanning window size, are very promising. However, the accuracy of thickness estimates were significantly reduced by the inability of the technique to distinguish between the tissue layer and large diameter nerve fibres apposed to the inner tissue boundary, which, accordingly, inflates the affected tissue thickness estimates. In addition, the tendency of the AR-SE algorithm to produce some thickness estimates of 0 µm, which is biologically impossible, raises the question of robustness of using the full-width-half-mean of the contiguous scatterer peak as an estimate of the base-width. Further studies on the relationship between the contiguous scatterer layer and epineurium thickness are needed.

In the CoD-SE technique the epineurium tissue layers were visually identified as a contiguous layer of scatterers, however, a method to isolate this layer and a quantitative method to analyse the thickness were not identified in the current study. Application of the CoD-SE technique to estimating epineurium thickness from the contiguous layer is expected to suffer from the same problem identified in the AR-SE technique: an inability to distinguish epineurium tissue from clusters of fibres adjacent to the internal tissue boundary. This inability appears to be an inherent shortfall in the application of scatterer diameter estimation techniques to distinguishing tissue layers.

Differences in fibre diameter distribution between fascicles were correctly identified by both the AR-SE and normalised CoD-SE techniques, however, the magnitude of this difference was not estimated accurately by either technique. In the AR-SE technique, we postulate that the estimate accuracy may be improved through the use of more biomimetic samples to populate the coefficient matrix within the algorithm. However, producing samples which mimic densely packed nerve fibres within the intra fascicle volume using materials with comparable refractive indices poses a challenge. Alternatively, spectral data gathered from fascicles of known fibre diameter distribution could be used to train the algorithm. Conversely, because the CoD-SE technique uses a Mie theory based model to predict the spectra, the challenge lies in locating suitable values for the refractive indices of the biological tissue and cellular layers. Furthermore, the myelinated fibre contains an internal scattering boundary between the myelin sheath and intracellular fluid of the axon which has not been considered in the current study.

In all techniques, inaccuracy was compounded by the low axial resolution of our OCT system (10 µm in air) relative to the lower bounds of epineurium tissue layer thickness (10 µm) present in some areas of each tissue sample. None of the techniques could identify the epineurium boundaries beneath adipose tissue or on the far side of the intra-fascicle volume. Therefore, application of the techniques appears to be limited in depth, perhaps to as little as the first $100 - 200$ µm, due to limitations of the techniques and to the highly scattering nature of adipose tissue and nerve fibres. This is in broad agreement with an observation by authors of the original ADRP technique [18], performed on an OCT system with 11 µm axial resolution in air, that the ADRP technique is unreliable at greater than 300 µm depth due to a decrease in the SNR. The authors of the original AR-SE technique, performed on an OCT system with 13 µm axial resolution in air, qualitatively implied the presence of three fascicles within an OCT image of a peripheral nerve at depths of up to 500 µm [19], although, as the authors point out, without any validation. Several 100's of µm depth penetration is sufficient to image the minor branches of the major nerves in humans, such as the pronator teres, flexor carpi radialis, and digital branches of the median nerve [9]. Furthermore, some medical applications, such neural tissue identification as an aid to surgery [11], do not require significant depth penetration, whereas others, such as monitoring of epineurium thickness in response implantation of nerve-cuff electrodes [34], or monitoring of myelination post crush injury [17], may garner useful information from the outer layers of the nerve alone. The principal advantage of OCT in extracting morphometric information from peripheral nerve is the wealth of information present in the OCT signal. As was the case in the current study, several signal processing techniques can be applied to a single set of OCT data, obtained from each nerve, in order to extract multiple morphometric details such as epineurium thickness (ADRP and AR-SE techniques), fibre diameter distribution (AR-SE and CoD techniques), and tissue classification (ADRP technique). Variation of morphometric details along the length of the nerve could then be easily be accomplished by using C-scans.

## 5 Conclusion

Three OCT signal analysis techniques were evaluated and improved upon in this study as a means to extract morphometric details of peripheral nerves. New methods of estimating the epineurium thickness were identified, and initial accuracy of results are promising. Further development of these techniques and the use of higher resolution OCT system are expected to improve the accuracy. Methods of quantifying the fibre diameter distribution were not successfully produced, however, factors that potentially improve the methods were identified. This study has characterised some of the abilities and limitations of OCT in extracting morphometric information of peripheral nerve and identified future research directions in this area.